\documentclass[aps,prl,twocolumn,showpacs,superscriptaddress,floatfix]{revtex4}
\usepackage{color,graphicx}
\begin{document}

  \author{S.\ Miyahara}
  \affiliation{
    Multiferroics Project (MF), ERATO,
    Japan Science and Technology Agency (JST), 
    Tokyo 113-8656, Japan}
  \author{K.\ Takubo}
  \affiliation{
    Department of Physics, Waseda University, 
    Tokyo 169-8555, Japan}
  \author{T.\ Suzuki}
  \affiliation{
    Department of Physics, Waseda University, 
    Tokyo 169-8555, Japan}
  \author{T.\ Katsufuji}
  \affiliation{
    Department of Physics, Waseda University, 
    Tokyo 169-8555, Japan}
  \affiliation{
    Kagami Memorial Laboratory for Material Science and Technology, 
    Waseda University, Tokyo 169-0051, Japan}
  \affiliation{
    PRESTO, Japan Science and Technology Agency, 
    Saitama 332-0012, Japan}
  \author{N.\ Furukawa}
  \affiliation{
    Multiferroics Project (MF), ERATO,
    Japan Science and Technology Agency (JST), 
    Tokyo 113-8656, Japan}
  \affiliation{
    Department of Physics and Mathematics, 
    Aoyama Gakuin University, 
    Kanagawa 229-8558, Japan}
  \title{
    Raman scattering due to a one-magnon excitation process
    in ${\rm MnV_2O_4}$}
  \date{\today}

  \begin{abstract}
    Unconventional peak structure in the Raman spectra due to magnon excitation
    at low temperature is observed 
    in spinel magnet ${\rm MnV_2O_4}$, 
    where a noncollinear spin state is realized by geometrical frustration. 
    We propose a new mechanism to induce such a 
    Raman scattering process due to a one-magnon excitation 
    of the noncollinear spin state.
    Novel features of the scattering 
    such as selection rules and peak position 
    observed experimentally in ${\rm MnV_2O_4}$  
    can be explained quite naturally by 
    considering the present one-magnon process. 
    We also discuss that such one-magnon process may exist in
    various materials with noncollinear spin structures.
  \end{abstract}


  \maketitle 
  Geometrically frustrated magnetic systems
  have been paid considerable attentions to 
  because of the novel magnetic 
  properties~\cite{ramirez99,kimura03,lee07,castelnovo08}.
  The frustration plays a role to suppress the conventional magnetic
  order, and stabilizes unusual ground states 
  with unusual magnetic excitations like
  macroscopically degenerate ground
  state in classical spin systems,
  a spin liquid ground state and spin gap excitations
  in quantum spin systems.
  As a typical example in three dimensional case,
  magnetism on a pyrochlore lattice has been 
  studied intensively more than 50 years~\cite{anderson56},
  and, novel features have still been discovered
  both experimentally and theoretically~\cite{ramirez99,castelnovo08}.

  One of such frustrated system 
  in three-dimensional system is a spinel vanadate 
  ${\rm MnV_2O_4}$~\cite{plumier87,plumier89,adachi05,suzuki07,garlea08,chung08},
  where a network of ${\rm V}^{3+}$ spins ($S_B = 1$)
  makes a pyrochlore lattice.
  In the ground state, ${\rm Mn}^{2+}$  spins $(S_A = 5/2)$
  and the  ${\rm V}^{3+}$ spins show a 
  noncollinear ferrimagnetic order due to the frustration.
  It was found that this compound reveals 
  the phase transitions at $T_{N}=58$ K.
  At $T_{N}$, the magnetic moment of the Mn and V sites align
  to the opposit direction.
  At the same temperature, a structural phase transition 
  from a cubic to a tetragonal occurs
  due to the orbital ordering on the V site~\cite{suzuki07,garlea08},
  where ferro-orbital ordering structure 
  along [$110$] and [$1\bar{1}0$] V chains,
  and antiferro-orbital ordering along $[001]$ directions are realized.
  Such an orbital structure partially relaxes the effects of 
  the geometrical frustration for V spins.
  As a result, the magnetic behaviors on V spins can be described 
  as the  weakly coupled antiferromagnetic chain along [$110$] 
  and [$1\bar{1}0$],
  and induce the magnetically ordered state.~\cite{tsunetsugu03}.

  In this Letter, we report Raman scattering results
  in ${\rm MnV_2O_4}$. At low temperatures ($T \ll T_N$),
  we observe a well-defined sharp peak corresponding 
  to the magnon mode. The peak appears even in parallel light 
  polarization, which can not induce the one-magnon scattering
  by conventional mechanisms~\cite{elliott63,moriya67}.
  Alternatively, we propose a novel mechanism such that,
  in a noncollinear spin structure, 
  Raman scattering process due to single magnon excitation 
  can also be generated for the parallel light polarization. 
  We demonstrate that the selection rule and the peak position
  of the Raman scattering observed in ${\rm MnV_2O_4}$ 
  are explained naturally  by the new mechanism. 

  Raman scattering was measured 
  on the cleaved (001) surface (in the cubic setting)
  of a ${\rm MnV_2O_4}$ single crystal, 
  which was grown by a floating-zone technique.
  The 514.5 nm laser line from an Ar ion laser was used 
  as the incident light, and
  the scattered light was collected 
  into a grating spectrometer with a CCD detector.
  The polarization directions, $a$ and $b$, are along the V-O bond, 
  and the $a^{\prime}$
  and $b^{\prime}$ are rotated by $45^\circ$ within the plane.
  Since the surface is determined in the cubic setting,
  a light polarization in the experiment may consist of two types of 
  polarization structures. Namely,
  ($a,b$) configuration can be ($x,y$) and/or ($z,x$),
  where ($x,y,z$) are defined in the tetragonal setting:
  $x$ and $y$-axes are along the longer V-O bond
  and $z$ is along shorter one.
  Hereafter, we distinguish the representation ($x,y,z$) 
  from ($a,b,c$) defined in the cubic structure. 
  Symmetry argument for the observed Raman modes was performed 
  in $D_{4h}$ space group, since we can not distinguish 
  $x$ ($y$) and $z$ directions at $T < T_N$.

\begin{figure}
    \begin{center}
      \includegraphics[width=0.95\columnwidth]{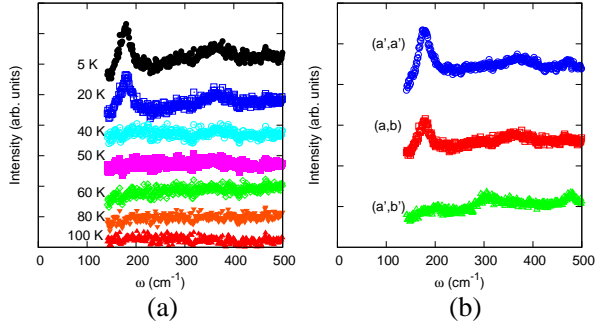} 
    \end{center}
    \caption{
      (Color online) (a) Temperature dependence of Raman spectra 
      for ($a,b$) light polarization.
      (b) Raman spectra for light polarizations
      ($a^{\prime},a^{\prime}$), ($a,b$), and
      ($a^{\prime},b^{\prime}$) at 5K.
    }
    \label{fig:Raman}
  \end{figure}

  The temperature dependence of Raman shift
  intensities of ${\rm MnV_2O_4}$ for $(a,b)$ light polarization 
  is shown in Fig.~\ref{fig:Raman} (a).
  At the lowest temperature $T = 5$K,
  a sharp peak around $180 {\rm cm^{-1}}$
  has been observed and the intensities of it increase by lowering
  the temperature below $T_N$.
  Light polarization dependence for
  $(a,b)$, $(a^{\prime},a^{\prime})$, and $(a^{\prime},b^{\prime})$
  at $T = 5$ K is shown in Fig.~\ref{fig:Raman} (b).
  The sharp peak has been observed for 
  both $(a,b)$ and  $(a^{\prime},a^{\prime})$ polarizations, but
  not for $(a^{\prime},b^{\prime})$.
  No new Raman-active phonon mode is expected by the
  structural phase transition from a cubic spinel to the
  tetragonal ($C_{4h}$) phase. 
  Energetically, the peak around 22 meV ($180$ ${\rm cm^{-1}}$) can be assigned
  to the single magnon excitation observed in
  the inelastic neutron scattering 
  experiment at ${\bf q} = (2, 0, 2)$~\cite{chung08}.
  The observed peak is a $B_{2g}$ symmetry mode
  in the space ground $D_{4h}$. 
  In this way, the peak is likely realized 
  through the magnetic excitation processes.

  However, the observed feature is unconventional 
  as a magnetic scattering as shown below.
  As the possible conventional processes, 
  we discuss 
  (i) one-magnon excitation and
  (ii) simultaneous two-magnon excitation, 
  which fail to explain the experimentally observed features.
  One-magnon scattering has been expected
  though the Raman tensor operator proportional to
  single spin term $S_i^\alpha$~\cite{moriya67}.  
  For the system where the inversion exist at the spin site,
  such a process can be active only for
  a cross light polarization. Thus, this synario is not
  consistent with the observation that the peak
  has been observed even in ($a^\prime,a^\prime$)  
  light configuration (see Fig.~\ref{fig:Raman} (b)).
  On the other hand, it is well known that
  the Raman tensor operator proportional to
  two spin term ${\bf S}_i \cdot {\bf S}_j$ can induce
  the scattering due to simultaneous two-magnon excitations
  in N\'{e}el ordered state~\cite{moriya67}.
  In this case, the 
  spectra of two-magnon scattering reflect the
  density of states of two-magnon processes and, thus,
  the spectrum consists of asymmetric broad peak. 
  In addition, the spectrum shape should depend on the light polarization.
  The observed peak in ${\rm MnV_2O_4}$ is likely a 
  symmetric single peak, whose shape is almost identical for ($a, b$) and 
  ($a^\prime,a^\prime$) configurations,
  and thus is unlikely two-magnon spectrum.

  To explain the origin of the observed peak,
  we introduce a new mechanism to induce a 
  Raman scattering process due to one-magnon 
  excitation in noncollinear structures of spins.
  Let us consider the Raman process induced 
  by a Heisenberg spin form tensor:
  \begin{equation}
    \hat{R} = \sum_{ij} \rho_{ij} ({\bf E}_{I}^{\omega} \cdot {\bf r}_{ij})
    ({\bf E}_{S}^{\omega} \cdot {\bf r}_{ij})
    \left({\bf S}_i \cdot {\bf S}_j \right),
    \label{eq:R2_dominant}
  \end{equation}
  where ${\bf r}_{ij}$ is the unit vector connecting 
  spin $i$ and $j$, and
  ${\bf E}_I^{\omega}$ and  ${\bf E}_S^{\omega}$ are the incident and 
  scattering light, respectively.
  The tensor (\ref{eq:R2_dominant}) can be driven by
  the perturbation calculation 
  in the Hubbard model~\cite{shastry90,vernay07} and 
  is dominant as a two spin 
  term~\cite{elliott63,moriya67,lemmens97,lemmens00,lemmens03}. 
  The coefficient $\rho_{ij}$ depends on the exchange coupling $J_{ij}$.
  In the perturbation, $\rho_{ij}$ is proportional 
  to $J_{ij}$~\cite{shastry90,vernay07}.
  In a collinear magnetic structure such as N\'{e}el ordered states,
  and also in a singlet gapped ground state,
  such a process can excite two magnon 
  and bound state of triplets excitations
  rather than single magnon and triplet excitations,
  {\it i.e.} excitation satisfies the 
  condition $\Delta S \equiv S_e - S_0= 0$
  where $S_0$ ($S_e$) is a spin number of the ground (excited) state. 
  However, in the noncollinear state, situation is dramatically changed
  so that one-magnon excitation by this process 
  can contribute to the Raman scattering process.

\begin{figure}
    \begin{center}
      \includegraphics[width=0.98\columnwidth]{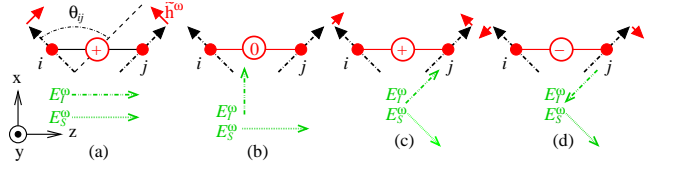}
    \end{center}
    \caption{
      (Color online) Effective fields $\tilde{h}^{\omega}$ and
      spin directions are shown  
      by small solid (red) and dash-dotted (black) arrows,
      respectively.
      The sign of $({\bf E}_{I}^{\omega} \cdot {\bf r}_{ij})
      ({\bf E}_{S}^{\omega} \cdot {\bf r}_{ij})$ on the bonds,
      which determine the direction of effective fields,
      are indicated.
      Light polarizations are (a) $(E_I^{\omega}, E_s^{\omega}) = (x, x)$, 
      (b) $(y, x)$, (c) $(x^{\prime}, -y^{\prime})$, and 
      (d) $(-x^{\prime}, -y^{\prime})$.
    }
    \label{fig:Raman-bond}
  \end{figure}

  To discuss the one-magnon scattering,
  let us rewrite the operator on the bond
  as $\hat{R}_{ij} = (\tilde{\bf h}_{i}^{\omega (j)} \cdot {\bf S}_i 
  + \tilde{\bf h}_{j}^{\omega (i)} \cdot {\bf S}_j)/2$ where 
  $\tilde{\bf h}_{i}^{\omega (j)} 
  = \rho_{ij} ({\bf E}_{I}^{\omega} \cdot {\bf r}_{ij})({\bf E}_{S}^{\omega} 
  \cdot {\bf r}_{ij}) {\bf S}_j$
  is a local effective field coupled to ${\bf S}_i$ and
  $\tilde{\bf h}^{\omega (i)}_{j}$ vice versa.
  The effective field can be approximated as
  \begin{equation}
    \tilde{\bf h}_{i}^{\omega (j)}\sim 
    \rho_{ij} ({\bf E}_{I}^{\omega} \cdot {\bf r}_{ij})
    ({\bf E}_{S}^{\omega} \cdot {\bf r}_{ij})
    \langle {\bf S}_j \rangle,
  \end{equation}
  where
  $\langle {\bf S}_j \rangle$ is an expectation value of the
  spin moment at site $j$ on the ground state.
  When such an effective field has a transverse 
  component with respect to the ground state 
  spin direction, it is possible to make
  $\Delta S = \pm 1$ excitation process, which 
  induces the one-magnon excitation.
  The magnitude of the transverse component is 
  represented by $\tilde{h}_{i \perp}^{\omega (j)} =
  \rho_{ij} ({\bf E}_{I}^{\omega} \cdot {\bf r}_{ij})
  ({\bf E}_{S}^{\omega} \cdot {\bf r}_{ij})
  S \sin \theta_{ij}$ as a function of the relative angle 
  $\theta_{ij}$ between ${\bf S}_i$ and ${\bf S}_j$. 
  Since the transverse component proportional to
  $\sin \theta_{ij}$, $\tilde{h}_{i \perp}^{\omega (j)}$ vanishes 
  for the collinear structure and becomes finite only for the
  noncollinear structure. Namely,
  $\Delta S = \pm 1$ excitation can be induced 
  when ${\bf S}_i$ and ${\bf S}_j$ are noncollinear.
  The effective field strongly depends on the light 
  polarization through the term $({\bf E}_{I}^{\omega} \cdot {\bf r}_{ij})
  ({\bf E}_{S}^{\omega} \cdot {\bf r}_{ij})$. As a result,  
  in the case that both $E_I^{\omega}$ and $E_S^{\omega}$ have a components 
  along the bond direction, the effective field 
  can exist (see Fig.~\ref{fig:Raman-bond}).
  Note that the direction of the effective fields can be reversed 
  by changing the direction of ${\bf E}_{I}^{\omega}$ or ${\bf E}_{S}^{\omega}$
  as shown in Figs.~\ref{fig:Raman-bond} (c) and (d).
  
  As a whole, the local effective field at site
  $\tilde{\bf h}_i^{\omega\,tot}$
  can be written as 
  \begin{equation}
    \tilde{\bf h}_i^{\omega\, tot} \equiv \sum_{j} \!^{\prime} \,
    \tilde{\bf h}_{i}^{\omega (j)} 
    = \sum_{j} \!^{\prime} \,
    \rho_{ij} ({\bf E}_I^{\omega}  \cdot {\bf r}_{ij})
    ({\bf E}_s^{\omega} \cdot {\bf r}_{ij})  \langle {\bf S}_j \rangle,
    \label{eq:effH}
  \end{equation}
  where the summation of $j$ is taken over all bonds connected 
  to $i$ site.
  When $\tilde{\bf h}_i^{\omega\,tot}$ has a transverse component,
  the Raman process (\ref{eq:R2_dominant}) couples to one-magnon excitation
  in ordered magnet.
  Once we know the ground state spin configuration,
  an activity of one-magnon Raman process and 
  an induced magnon mode can be
  checked easily from Eq.(\ref{eq:effH}).
  The selection rule can be understood from the 
  light polarization dependence of the effective fields (\ref{eq:effH})
  as well.
  The Raman shift energy corresponds to
  the one-magnon excitation energy which can be evaluated   
  by applying a conventional linear spin wave theory.
  Here excited magnon wavenumber ${\bf q}$ can be obtained
  from the Fourier transforms of the effective 
  field $\tilde{\bf h}_i^{\omega\,tot}$.
  Note that, when we consider the magnon dispersion 
  in a reduced Bourillouan zone of the magnetically ordered state,
  such a wave number ${\bf q}$ must exist at $\Gamma$ points.
  
\begin{figure}
    \begin{center}
      \includegraphics[width=1.0\columnwidth]{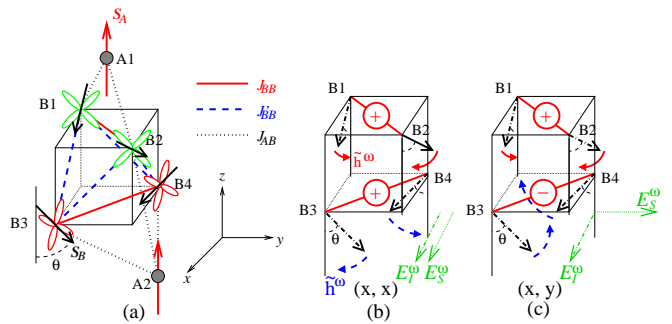}
    \end{center}
    \caption{
      (Color online) (a) Magnetic structures and exchange interactions
      $J_{AB}$, $J_{BB}$, and $J_{BB}^\prime$  for ${\rm MnV_2O_4}$. 
      Ground state spin structures are shown.
      Orbital order structure on $B$-sites
      are also indicated:  $d_{yz}$ on $B1$ and $B2$ sites and
      $d_{zx}$ on $B3$ and $B4$ sites.
      (b) Effective fields for $V$ spins 
      with light polarizations $(x,x)$ and
      (c) $(x,y)$. The solid (dashed) arrows in $zx$ ($yz$) plane.
      The dash-dotted arrows indicates the spin direction
      on the ground states.
    }
    \label{fig:model}
  \end{figure}

  Let us now bring the argument back to the topics on  
  ${\rm MnV_2O_4}$.  As shown in Fig.~\ref{fig:model} (a), 
  a noncollinear ground state structure has been observed in
  the neutron scattering experiment~\cite{garlea08}. 
  In this compound, the strongest super exchange interaction 
  exists on the bond along [$110$] and [$1\bar{1}0$] directions
  between V sites ($J_{BB}$ bond in Fig.~\ref{fig:model} (a)).
  Thus, we consider the Raman process on this bond.
  From the figure,
  we can easily obtain that one-magnon Raman process is 
  active in this compound, and two types of magnon modes
  can be induced for $(x,x)$ and $(x,y)$ light polarizations.
  Each mode can be assigned to $A_{1g}$ and $B_{2g}$ symmetry mode 
  respectively in the space group $D_{4h}$.
  Effective fields of each magnon mode are
  shown in Figs.~\ref{fig:model} (b) and (c).
  Since ($x^\prime, x^\prime$) polarization contains
  $A_{1g} + B_{2g}$  symmetry, both magnon modes
  produce the Raman scattering.
  On the other hand, ($x^\prime, y^\prime$) polarization 
  does not couple to the one-magnon scattering
  since one of the components is orthogonal 
  to the bond direction as in Fig.~\ref{fig:Raman-bond} (b).
  From the symmetry of light polarization, 
  the experimentally observed $B_{2g}$ mode at $180 {\rm cm}^{-1}$
  can be assigned to the one-magnon mode in Fig.~\ref{fig:model} (c)
  and the peak structure in Raman spectra for
  $(a,b)$ light polarization is likely realized due to
  $(x,y)$ configuration rather than $(z,x)$.
  The selection rule for the present mechanism predicts that
  the other mode in  Fig.~\ref{fig:model} (b) exists at lower energy.
 
  To confirm the peak position of each magnon mode,
  we consider the magnetic features in 
  the spin Hamiltonian for ${\rm MnV_2O_4}$~\cite{chung08}:
  \begin{equation}
    {\cal H} = \sum_{n.n.} \tilde{J}_{ij} \tilde{\bf S}_{i} \cdot \tilde{\bf S}_j
    + \sum_{i \alpha} \tilde{D}_i^\alpha (\tilde{S}_i^\alpha)^2,
    \label{eq:H}
  \end{equation}
  where ${\bf \tilde{S}}_i$ is an $S = 1$ spin operator and
  interactions are defined as $\tilde{J}_{ij} = J_{ij} S_i S_j$ 
  and $\tilde{\bf D} = {\bf D} (S_i)^2$. 
  The nearest neighbor interactions between $A$ and $B$ 
  are defined as $\tilde{J}_{AB}$ and those between $B$ ions
  $\tilde{J}_{BB}$ and $\tilde{J}_{BB}^{\prime}$ as
  shown in Fig.~\ref{fig:model} (a).
  The single ion anisotropy term at each spin site 
  is also introduced as $\tilde{\bf D}_A$ and $\tilde{\bf D}_B$, respectively.
  We assume the spin structures estimated
  from the neutron scattering experiment~\cite{garlea08}:
  $\tilde{S}_{A1} = (0, 0, 1)$, $S_{A2} = (0, 0, 1)$,
  $\tilde{S}_{B1} = (\sin \theta, 0, -\cos \theta)$,
  $\tilde{S}_{B2} = (-\sin \theta, 0, -\cos \theta)$,
  $\tilde{S}_{B3} = (0, \sin \theta, -\cos \theta)$, and
  $\tilde{S}_{B4} = (0, -\sin \theta, -\cos \theta)$.
  The angle $\theta$ is given by  $\cos\theta = 6 \tilde{J}_{AB} / 
  (4 \tilde{J}_{BB} + 4 \tilde{J}_{BB}^\prime 
  + 2 \tilde{D}_B^z  - \tilde{D}_B^x - \tilde{D}_B^y)$
  to minimize the classical ground state energy.
  The orbital ordering 
  of $d_{yz}$ and $d_{zx}$ on $B$-sites, {\it i.e.}, V-sites,
  observed in Refs.~\onlinecite{suzuki07} and \onlinecite{garlea08} 
  is also shown in the figure.

\begin{figure}
    \begin{center}
      \includegraphics[width=0.95\columnwidth]{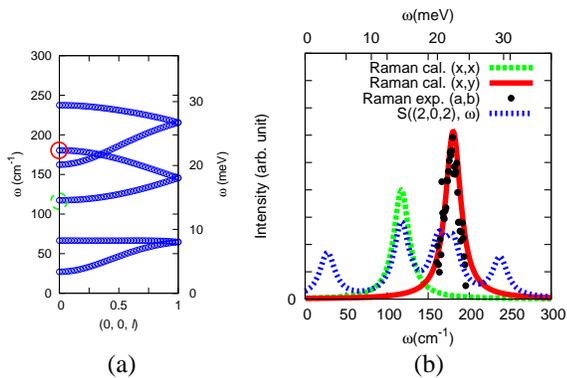}
    \end{center}
    \caption{
      (Color online) (a) Magnon dispersion for the present model 
      along (0,0,l) direction.
      Raman active energy for ($x,y$) (($x,x$)) light polarization
      is indicated by red solid (green dashed) circles. 
      (b) Raman shifts intensities for ($x,x$) 
      and ($x,y$) light polarizations, and  
      $S({\bf q}, \omega)$ at ${\bf q} = (2, 0, 2)$.
      The delta function is replaced by the Lorentzian with
      width $\epsilon/\tilde{J}_{BB} = 0.1$.
      The peak observed experimentally for ($a,b$) light 
      polarizations at 5K 
      (experimental results between 160 ${\rm cm^{-1}}$ and 
      195 ${\rm cm^{-1}}$) is also shown.
    }
    \label{fig:Sq}
  \end{figure}

  To reproduce the magnetic behaviors in ${\rm MnV_2O_4}$,
  we adopt the interactions:
  $\tilde{J}_{BB} = 14.0$ meV (113 ${\rm cm}^{-1}$),
  $\tilde{J}_{AB}/\tilde{J}_{BB} = 0.18$,
  $\tilde{J}_{BB}^{\prime}/\tilde{J}_{BB} = -0.15$,
  $\tilde{D}_A^z/J_{BB} = -0.01$ for $A$ site ions,
  $\tilde{D}_B^{\beta} = -0.15$ ($\beta =x$ for $B1$ and $B2$ sites
  $y$ for $B3$ and $B4$), and
  $\tilde{D}_B^{z} = 0.2$ for $B$ site ions.
  The other components of  $\tilde{\bf D}$ are take to be zero.
  Calculated dynamical structure factor 
  $S({\bf q}, \omega)$ along ($4-l$, 0, $l$) and (2, 0, $l$) 
  by the linear spin wave theory
  is reasonably consistent with experimental data observed in neutron 
  scattering experiments~\cite{chung08}. 
  As a typical case,
  the structure factor $S({\bf q}, \omega)$ at ${\bf q} = (2, 0, 2$) 
  is shown in Fig.~\ref{fig:Sq}
  (strong intensity between 20 and 25 meV
  of $\hbar \omega S({\bf q}, \omega)$ in Ref.~\onlinecite{chung08}
  corresponds to the two peaks structure in this energy range).
  $\tilde{J}_{BB}^\prime$ is taken as a
  ferromagnetic interaction, which is reasonable for the
  antiferro-orbital ordering state~\cite{tsunetsugu03,memo-chung}.
  In our model, the canted angle $\theta$ 
  for the V moment is about $75^{\circ}$, 
  which is reasonably consistent with
  the value $65^{\circ}$ estimated from neutron scattering 
  experiments~\cite{garlea08}.
  Note that the magnetic excitations obtained in the given parameters
  give the much better agreement with the experimental results
  than those in the parameters with $\theta = 65^{\circ}$.

  Let us consider the magnetic excitation 
  at $\Gamma$ points (Magnon dispersion along (0,0,$l$) is shown
  in Fig.~\ref{fig:Sq} (a)).
  Lowest two energy modes (30 and 80 ${\rm cm}^{-1}$) 
  are due to oscillations of $S_A$ spins and
  the other four modes are mainly caused by vibrations of $S_B$ spins 
  as observed in related spinel ${\rm Mn_3O_4}$~\cite{chung08}.
  The magnon excitation around $180$ ${\rm cm}^{-1}$
  has the magnon mode in Fig.~\ref{fig:model} (c), which is
  in agreement with our assignment from the symmetry of light polarizations.
  The magnon mode for ($x,x$) (the mode in Fig.~\ref{fig:model} (b))
  is expected to exist at lower frequency around $120$ ${\rm cm^{-1}}$.
  Experimental observation at lower frequencies will be the
  test of the validity of our theory.
  ($x^\prime,x^\prime$) light polarization consist of
  both ($x,x$) and ($x,y$) features.
  Thus, for this configuration, two peaks are expected, and
  the higher energy peak is consistent with the experimental observation.
  Theoretically, there is no peak due to one magnon process
  for ($x^\prime,y^\prime$) configuration
  as the experimental observation
  in the energy range $\omega \gtrsim 150 {\rm cm}^{-1}$.
  The peaks in Raman spectra due to one-magnon process calculated
  for ($x,x$) and ($x,y$) light polarizations,
  and the experimentally observed peak for ($a,b$),
  which is likely produced by ($x,y$) light polarization,
  are shown in Fig.~\ref{fig:Sq} (b). 
  In our theory, there is no magnetic scattering 
  due to one magnon process for light polarizations
  on the $yz$ and $zx$ plane except for $(x,x)$ and $(y,y)$.
  Thus, experiments on the surfaces fixed in the tetragonal
  setting also give the crucial test to our theory.

  To summarize, we propose a new mechanism 
  to induce one-magnon Raman scattering
  process in noncollinear magnets. 
  Such a one-magnon process for a certain light polarization 
  can detect a single one-magnon mode at $\Gamma$ point,
  where several modes are folded within the reduced zone.
  One of the advantages of the Raman scattering is that it is sensitive
  to the symmetry of the
  magnon mode so that, together with the selection rules, it clarifies
  microscopic details of the magnon excitations.
  On the other hand, neutron scattering experiments detect the magnon
  dispersions throughout the Brillouine zone with less sensitivity to
  the symmetry.
  Thus, both methods play complementary roles to investigate the
  magnetic excitations in noncollinear magnets.
  In the present case, we indeed obtain the spin exchange couplings
  which matches the orbital-order patterns of ${\rm MnV_2O_4}$.

  One-magnon Raman scattering process observed in 
  ${\rm MnV_2O_4}$ can also exist in a wide range of magnetic materials, 
  especially in frustrated magnets 
  with noncollinear spin structures.
  For example, such Raman processes are likely observable 
  in two-in two-out structures realized in the spin ice 
  materials~\cite{harris97,ramirez99,bramwell01} and 
  parasite ferromagnetic states as observed in
  $\alpha {\rm Fe_2O_3}$ above Morin temperature~\cite{morin50}.
  The possibility in the other materials can be 
  discussed by the same procedure in ${\rm MnV_2O_4}$.

  In this Letter, we restrict the tensor 
  operator in Eq.~(\ref{eq:R2_dominant}).
  Even in the case of reduced symmetries, {\it e.g.}
  the effects of the polarization perpendicular to the bond 
  can not be neglected, we can treat the process 
  using a general form for Raman operator in Ref.~\onlinecite{moriya67}
  in a similar fashion.
  
  We thank R. Kubota, N. Kida, T. Arima, R. Shimano,
  Y. Segawa, and Y. Tokura for fruitful discussion.
  This work is in part supported by Grant-In-Aids for Scientific 
  Research from the Ministry of Education, Culture, 
  Sports, Science and Technology (MEXT) Japan.

\bibliographystyle{apsrev}
\bibliography{Raman_short}

\begin{thebibliography}{23}
\bibitem[{mem()}]{memo-chung}
\bibinfo{note}{In Ref.~\cite{chung08}, the interaction $\tilde{J}_{BB}^\prime$
  has been estimated to be antiferromagnetic under the assumption of the
  ferro-orbital ordering. Note that the work has been done prior to the
  experimental determination of magnetic and orbital
  structures~\cite{garlea08}.}
\end{thebibliography}

\end{document}